\documentstyle[twoside,fleqn,espcrc2,epsf,rotate]{article}

\newcommand{\AmS}{{\protect\the\textfont2
  A\kern-.1667em\lower.5ex\hbox{M}\kern-.125emS}}

\newcommand{\plus}{\makebox[15pt][c]{$+$}}
\newcommand{\minus}{\makebox[15pt][c]{$-$}}
\newcommand{\er}[2]{\raisebox{0.08em}{\scriptsize {$\;\begin{array}{@{}l@{}}
                          \plus\makebox[0.90em][r]{#1} \\[-0.12em]
                          \minus\makebox[0.90em][r]{#2}
                        \end{array}$}}}
\newcommand{\ers}[2]{\raisebox{0.08em}{\scriptsize {$\;\begin{array}{@{}l@{}}
                          \plus\makebox[0.55em][r]{#1} \\[-0.12em]
                          \minus\makebox[0.55em][r]{#2}
                        \end{array}$}}}

\def\uplw{$\mbox{}$\er{up}{lw}}

\title{
\vspace*{-15pt}
{\normalsize November 1993 \hfill UTHEP-267}\\
High statistics calculations of quenched QCD spectrum
  using various quark
  sources\thanks{Talk presented by T.Yoshi\'e at Lattice '93}}

\author{QCDPAX collaboration: \\
Y.~Iwasaki\address{Institute of Physics, University of Tsukuba,
                   Ibaraki 305, Japan},
K.~Kanaya$\mbox{}^{\rm \ a}$,
S.~Sakai\address{Faculty of Education, Yamagata University,
                 Yamagata 990, Japan},
T.~Yoshi\'e$\mbox{}^{\rm \ a}$,
T.~Hoshino\address{Institute of Engineering Mechanics, University of Tsukuba,
                   Ibaraki 305, Japan}
and
T.~Shirakawa$\mbox{}^{\rm \ c}$
}

\begin{document}

\begin{abstract}
We present the results for the hadron spectrum calculated on
400 configurations using point source, wall source and 8-cubic sources,
in quenched QCD with Wilson fermions at $\beta=6.0$ and $K=0.155$
on a $24^3 \times 54$ lattice.
The results for the ground state masses obtained with
three types of quark sources agree well with each other.
Masses of the first excited states appear consistent with experimental
values within large errors.
\end{abstract}

\maketitle

Recent high statistics simulations of quenched QCD have
produced high quality data for the masses of light
hadrons\cite{QCDPAXLattice92,Ape600,Ape630,GF11Mass,UKQCD620}.
However, mass results obtained with various types of quark sources
often do not agree with each other\cite{LANLsmear,QCDPAXLattice92,Ape600}.
Masses of the first excited states also often appear unexpectedly
heavy\cite{Ape600,Ape630,UKQCDMatrixCorr}.

In order to resolve these problems, the QCDPAX Collaboration
has carried out a high statistics quenched calculation
with Wilson quarks at $\beta$= 6.0 on a $24^3 \times 54$ lattice.
Taking the last configuration of our previous work\cite{QCDPAXLattice92}
as a starting configuration,
we have generated new 400 configurations which are separated
by 1000 heat bath sweeps.
We choose $K = 0.155$:
This $K$ corresponds approximately to the strange quark mass,
because the $m_{\pi}/m_{\rho}$ appears to be 0.7.

We use three types of quark sources: point source, wall source and
8-origin cubic sources of a size $7^3$
(with a random $Z(3)$ element on each cube).
We call the third source multi-cube source.
Hereafter we abbreviate them to p-, w- and m- sources, respectively.
Point sinks are used for all sources.
Configurations are fixed to Coulomb gauge
before quark propagators are solved for the w- and the m- sources.
Although we have 600 samples for the p-source in total,
we report the results obtained by using the new 400 configurations,
in order to compare results with three types of sources
on an equal footing.

\begin{table*}[hbt]
\setlength{\tabcolsep}{0.5pc}
\newlength{\digitwidth} \settowidth{\digitwidth}{\rm 0}
\catcode`?=\active \def?{\kern\digitwidth}
\caption{Results for the ground state masses. In parentheses are errors
estimated by the jack-knife method.
Errors given in the form \protect\uplw \,\,
are for the fitting dependent upper/lower bound.}
\label{ground}
  \begin{tabular*}{\textwidth}{@{}l@{\extracolsep{\fill}}crcrcr}
  \hline
        &  \multicolumn{2}{c}{Point Source}
        &  \multicolumn{2}{c}{Wall Source}
        &  \multicolumn{2}{c}{Multi-Cube Source} \\
\cline{2-3} \cline{4-5} \cline{6-7}
        &  $t_{min}-t_{max}$  & mass(er)\er{up}{lw}
        &  $t_{min}-t_{max}$  & mass(er)\er{up}{lw}
        &  $t_{min}-t_{max}$  & mass(er)\er{up}{lw} \\
  \hline
   $\pi$      & $13-27$ & 0.2960(?8)\er{9}{20}        
              & $11-27$ & 0.2964(?6)\er{12}{5}        
              & $12-27$ & 0.2969(?6)\er{6}{13} \\     
   $\rho$     & $14-20$ & 0.4201(29)\er{25}{36}      
              & $13-20$ & 0.4228(19)\er{8}{15}       
              & $11-20$ & 0.4249(19)\er{5}{33} \\    
   $N$        & $13-24$ & 0.6403(50)\er{0}{161}      
              & $13-18$ & 0.6307(39)\er{44}{10}      
              & $14-18$ & 0.6208(72)\er{30}{23} \\   
   $\Delta$   & $16-23$ & 0.7054(95)\er{9}{171}      
              & $11-18$ & 0.7008(57)\er{14}{59}	     
              & $14-24$ & 0.7128(191)\er{13}{345} \\ 
   \hline
  \end{tabular*}
\end{table*}

Masses of the ground states are determined by global fits to data for
$[t_{min},t_{max}]$ for which the $\chi^2/$dof of the fits are less
than 1.5.
We take account of time correlations in the fits.
Errors are estimated by single elimination jack-knife procedure.
We have checked that errors do not change significantly
even if we change the bin size in the jack-knife error analyses.
In addition, we have estimated errors which depend on
the choice of fitting ranges.
Here we define the fitting-dependent upper (lower) bound by the
maximum (minimum) of fit results to data for any region
$[t_0,t_1] \subset [t_{min},t_{max}]$ ($t_1-t_0 > 2$).
Detail of our fitting procedure will be reported elsewhere.

Results for the ground state masses are reproduced in Table \ref{ground}
and are summarized as follows.
1) Long plateau in effective masses can be seen
for all particles and all sources.
Length of the plateau ($t_{max}-t_{min}+1$) is typically 15 for $\pi$,
8 for $\rho$, 6 for nucleon and 8 for $\Delta$.
2) Statistical and fitting-dependent errors are typically of order 2 \%.
3) Mass results obtained with three types of quark sources coincide within
0.3 \% for $\pi$, 1 \% for $\rho$, 4 \% for nucleon and 2 \% for $\Delta$.
They agree within errors.
(See also Figs. \ref{rho-mass} and \ref{del-mass}.)
It may be worth while emphasizing that
the $\Delta$ mass obtained with the p-source agrees with that with
the w-source with small errors.
Note that the error for the $\Delta$ mass is the largest for the m-source.
The agreement of the results obtained with various types of sources
implies that they have little contamination from
excited states.
(See also the contributions by Ape group\cite{ApeLat93}
and R.~Gupta\cite{LANLLat93}.)

\newbox\B
\epsfysize=200pt
\setbox\B=\vbox{\epsfbox{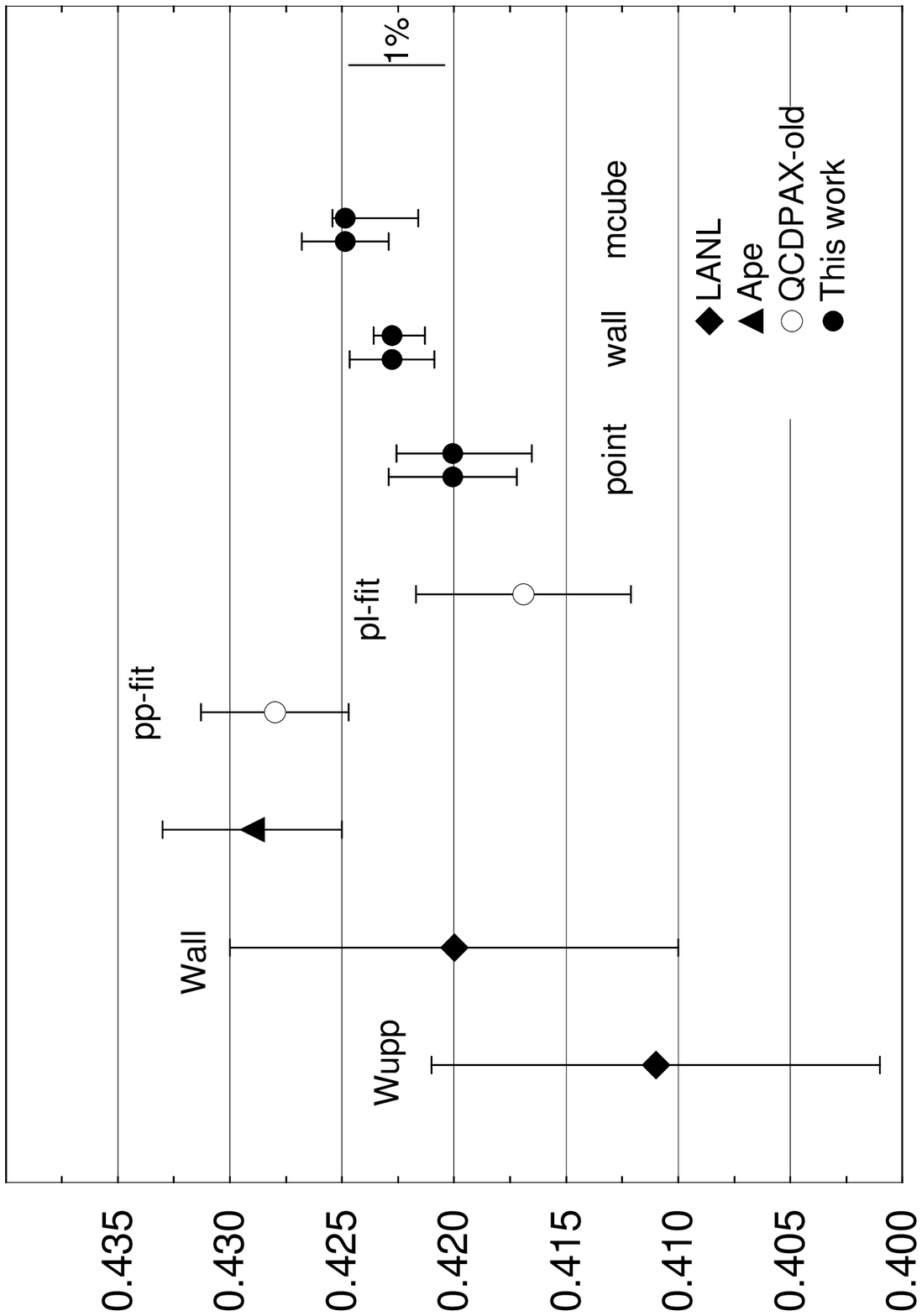}}

\begin{figure}[htbp]
\begin{center}
\leavevmode
  \makeatletter
  \@rotr\B
  \makeatother
\end{center}
\caption{$\rho$ meson masses obtained with various types of quark sources.
Two error bars are drawn for our new results (filled circles).
On the left are for errors estimated by the jack-knife method.
The right ones are for fitting dependent upper/lower bounds.
Other data are from LANL group\protect\cite{LANLsmear} (diamonds),
Ape Collab.\protect\cite{Ape600} (triangle)
and our previous work\protect\cite{QCDPAXLattice92} (open circles).}
\label{rho-mass}
\end{figure}

\newbox\C
\epsfysize=200pt
\setbox\C=\vbox{\epsfbox{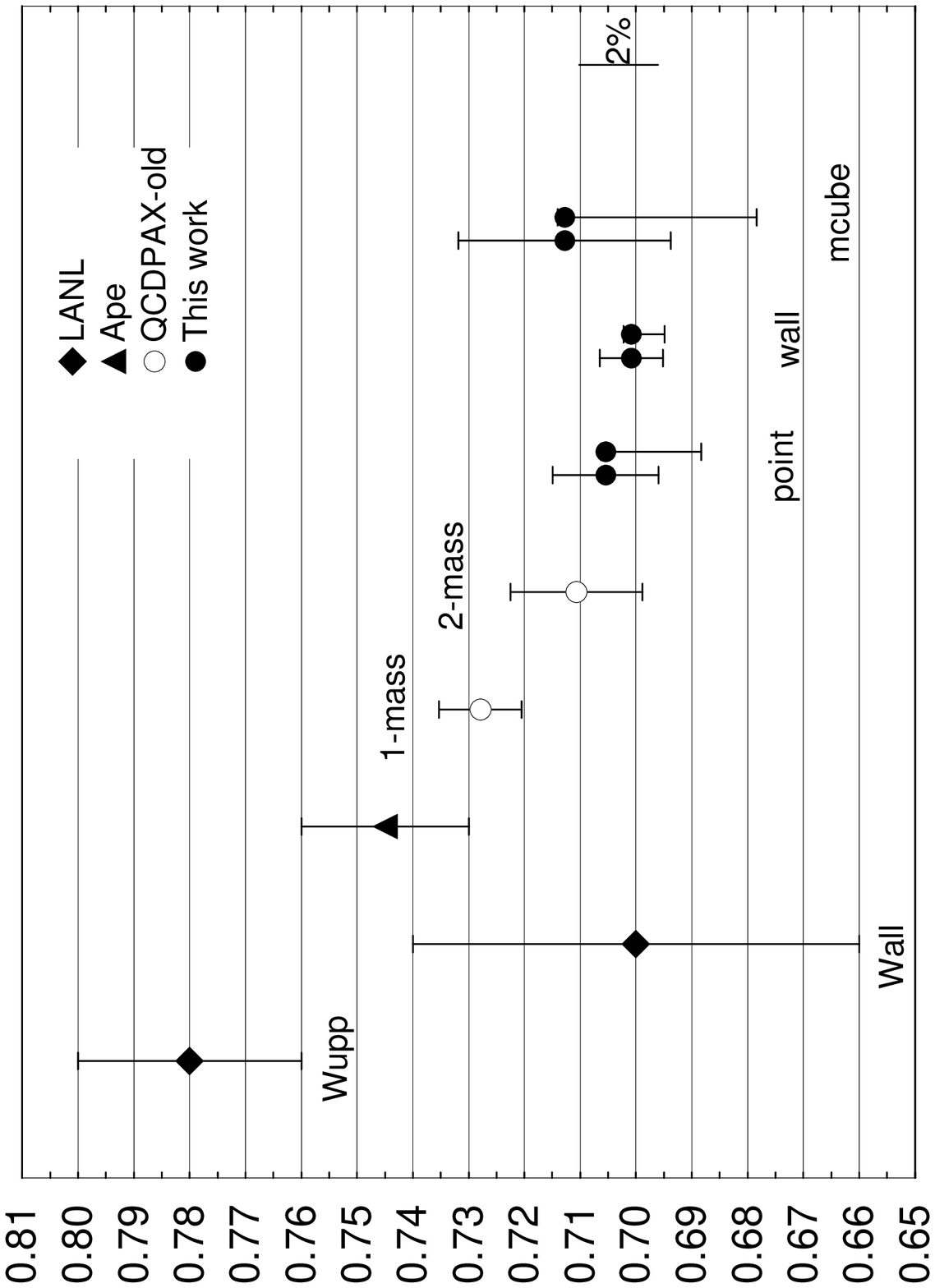}}

\begin{figure}[htbp]
\begin{center}
\leavevmode
  \makeatletter
  \@rotr\C
  \makeatother
\end{center}
\caption{The same as fig. \protect\ref{rho-mass} but for $\Delta$.}
\label{del-mass}
\end{figure}

The results for hadron masses are consistent with our previous ones.
In ref.\cite{QCDPAXLattice92}, we have made two different fits
(plateau fit and pre-plateau fit) to our old $\rho$ meson propagator and
obtained $m_{\rho}$= 0.4169(48) and 0.4280(33) for each fit, respectively.
We have taken the plateau fit to be correct.
Our new results are consistent with the value obtained by the plateau fit.
(See fig. \ref{rho-mass}.)
For baryons, new results are consistent with the old ones which are obtained
by two-mass fits; $m_{N}$= 0.635(11) and $m_{\Delta}$= 0.711(12).
Note that in our previous work the errors of baryon propagators
for the p-source are large at high time slices
and consequently the one mass fit gave a larger value than the two mass fit.
We have taken the result by the two mass fit as correct.
In this work the errors have been reduced significantly due to high
statistics.
Therefore the results of one mass fits are smaller than our
previous ones ($m_{N}$= 0.645(6) and $m_{\Delta}$= 0.728(7))
and are consistent with our previous results of
two mass fits.

Results for the mass ratios $m_{N}/m_{\rho}$ and $m_{\Delta}/m_{\rho}$
are consistent with values estimated by phenomenological
mass formulae\cite{Ono}:
$m_{N}/m_{\rho}$= 1.524\ers{13}{47}, 1.492\ers{16}{9}, 1.461\ers{18}{16}
and $m_{\Delta}/m_{\rho}$= 1.679\ers{24}{51}, 1.658\ers{13}{17},
1.678\ers{44}{83} \, for the p-, w- and m- sources, respectively.
We have estimated two kinds of errors.
One is the error obtained by the jack-knife method for the mass ratios.
The other is the fitting dependent error estimated by
dividing the fitting dependent upper (lower) bound for the numerator
by the fitting dependent lower (upper) bound for the denominator.
We have quoted the larger one.

Our next question is whether the masses of the first excited states
appear consistent with experiment.
We find that for each particle
there exist two mass fits with reasonable
$\chi^2$'s which give the ground state masses consistent
with one mass fits.
For these fits, almost all the results for the first excited state masses
are consistent with experiment within large one standard deviation.

Result for the $\rho$ meson is summarized as follows.
For the p-source, the fit with $t_{min}=7$
gives $m_{\rho}$ = 0.4184(34) and $m_{\rho}'$ = 0.836(63)
with $\chi^2/$dof $\sim 1.7$.
For the w-source,
all the fits with $t_{min}=4 \sim 7$ give $\chi^2/$dof $\sim 1$
and $m_{\rho}' \sim $ 0.7.
We quote the result from the fit with $t_{min}=6$:
$m_{\rho}$ = 0.4240(30), $m_{\rho}'$=0.72(11).
The results for $m_{\rho}'$ are consistent with the experimental value
$m_{\rho}' \sim 0.7$.
(This value is obtained assuming that the $K$ used
corresponds exactly to the mass of the strange quark and
using the experimental values for $\phi$ meson masses;
$\phi(1020)$, $\phi(1680)$.)
However, for the m-source,
there are only fits which are marginally consistent with $m_{\rho}' \sim 0.7$
with large errors, for example, 0.96(33) with $t_{min}=8$.

For nucleon, the fit with $t_{min}$= 8 (7, 6) for the p- (w-, m-) source
gives, with $\chi^2/$dof =1.1 (0.8, 1.2),
$m_{N}$= 0.6302(77) (0.6346(79), 0.6253(128))
which is consistent with the one-mass fit result.
The results for the $m_{N}'$ agree with each other with large errors:
$m_{N}'$ = 0.99(13), 0.93(13) and 1.01(7), respectively.
We estimate $m_{N}'= 0.8 \sim 1.1$ from these fits.
This value is smaller than that reported by Ape group (1.38(2))\cite{Ape600}.
We estimate the experimental value of the $m_{N}'$ to be 0.84
assuming that $m_{N}'- m_{N}$ at the strange quark mass is the same
as that at $u/d$ quark mass (500 MeV).

Two mass fits are more delicate for $\Delta$.
With the constraint of reasonable $\chi^2$,
only the fit with $t_{min}=9$ for the p-source gives
the result for the $m_{\Delta}$ which
is consistent with that obtained by one-mass fit.
This fit gives $m_{\Delta}$ = 0.695(14) and $m_{\Delta}'$= 1.02(11)
with $\chi^2/$dof=1.1.
Experimental value for the $m_{\Delta}'$ is 0.9 (from $\Omega(2250)^{-}$).

The numerical calculations have been performed with QCDPAX,
a parallel computer developed at University of Tsukuba.
This project is supported by the Grant-in-Aid
of Ministry of Education, Science and Culture
(No.62060001,No.02402003 and No.04NP0601)
and University of Tsukuba Project Research in 1993.
One of the authors (T.Y.) would like to thank Yamada Science Foundation
for financial support.


\begin{thebibliography}{9}

\bibitem{QCDPAXLattice92}
QCDPAX Collaboration (Y.~Iwasaki {\it et al.}), Nucl.~Phys.~B
(Proc.~Suppl.) 30 (1993) 397.

\bibitem{Ape600}
Ape Collaboration (S.~Cabasino {\it et al.}), Phys.~Lett. B258 (1991) 195.

\bibitem{Ape630}
M.~Guagnelli {\it et al.}, Nucl.~Phys. B378 (1992) 616.

\bibitem{GF11Mass}
F.~Butler {\it et al.}, Phys.~Rev.~Lett. 70 (1993) 2849.

\bibitem{UKQCD620}
UKQCD Collaboration (C.~R.~Allton {\it et al.}), Edinburgh preprint 93/524.

\bibitem{LANLsmear}
D.~Daniel {\it et al.}, Phys.~Rev. D46 (1992) 3130.

\bibitem{UKQCDMatrixCorr}
UKQCD Collaboration (C.~R.~Allton {\it et al.}),
Phys.~Rev.~D 47 (1993) 5128.

\bibitem{ApeLat93}
Ape Collaboration, presented by F.~Rapuano, these proceedings.

\bibitem{LANLLat93}
R.~Gupta {\it et al.}, these proceedings.

\bibitem{Ono}
S.~Ono, Phys.~Rev.~D 17 (1978) 888.

\end{thebibliography}
\end{document}